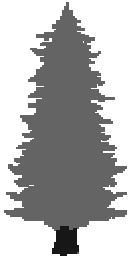

**2005 ALCPG & ILC Workshops – Snowmass, U.S.A.**

# General Thoughts about Tracking for a Linear Collider Detector

Bruce A. Schumm
*Santa Cruz Institute for Particle Physics, Santa Cruz, CA 95064, USA*

Basic scaling laws are used to compare the relative capabilities of gaseous and solid-state tracking. Conclusions are drawn in the context of physics processes expected to be at the focus of studies at the International Linear Collider

Current International Linear Collider (ILC) detector design concepts present two central tracking options: a large-volume TPC with limited contribution from silicon microstrip tracking at low radius (LDC, GLD), and a fully-silicon tracker that employs only microstrip tracking (SiD). Common knowledge seems to suggest that microstrip tracking, while precise, is inefficient with regard to the amount of information obtained per radiation length of tracking material, and that the precision associated with microstrip tracking leads to demands on alignment and calibration that threaten to compromise its performance. The purpose of this paper is to explore issues such as these with general scaling laws, and to draw conclusions from the scaling-law arguments in the context of expected ILC physics signals.

Indeed, the physics context itself has evolved somewhat over the past 12 months. Recent studies done at SLAC [1], the University of Michigan [2], and UC Santa Cruz [3] suggest that the measurement of LC benchmark physics signals such as Higgs boson properties and slepton masses can benefit from momentum reconstruction precision better than that expected for current design concepts. This is somewhat counter to prior thinking, for which it was felt that the precision of the current design concepts matched the intrinsic capabilities of the machine. The SLAC, Michigan, and Santa Cruz studies found this to be the case only for the warm design that is no longer under consideration, due to the relatively large (~1%) energy spread of the accelerator beams. For the selected superconducting technology, the smaller (~0.1%) beam energy spread appears to make the collider an intrinsically more precise instrument, thereby calling for a more precise detector.

Since the 1996 introduction of the notion of a LC detector tracker composed entirely of silicon microstrip sensors [4], there has been an ongoing discussion of the relative merits of gaseous (TPC-based) versus solid-state (silicon microstrip) tracking. With its surfeit of three-dimensional space points, TPC's are thought to be excellent candidates for the efficient reconstruction of charged particle tracks in dense jet environments. General wisdom seems to suggest also that, due to the gas gain of proposed readout systems, gaseous tracking is a relatively efficient way to extract spatial information from traversing particles, leading to a small contribution from multiple Coulomb scattering, and correspondingly good performance at low momentum. Solid-state tracking, on the other hand, has a reputation for yielding precise space points, and is ideally suited for achieving excellent precision over a the constrained redial lever

**ALCPG1312**

arm provided by a compact detector design. Because such detector designs rely more on precision, and less on radial lever arm, than more voluminous gaseous designs, calibration of all-silicon tracking detectors is thought by some to be a particularly crucial issue.

In fact, though, there is no intrinsic aspect of solid-state tracking that limits its use to compact detector designs. While the consideration of such designs certainly benefit from the precision offered by solid-state tracking, solid-state tracking systems can be built economically on any scale relevant to the study of ILC physics. In fact, the choice between compact, high-field and more spacious but lower-field tracking designs should more appropriately be driven by the need to minimize the beampipe radius through the constraint on the radial extent of electron/positron pairs that is related to the strength of the magnetic field. Thus, in this paper, I will consider the relative performance of gaseous and solid-state tracking systems that instrument a common volume, which I have chosen to be that of the gaseous-tracking based LDC design, which is essentially the same as that of the former TESLA detector design.

While studies are underway to discern the minimum number of solid-state tracking layers necessary to do ILC physics, few would argue with the contention that a large-volume TPC is a robust tracking system. Some question remains as to the effect of the relatively weak TPC track-separation resolution on the tracking efficiency within dense jets; however, this is unlikely to be a significant issue once mature reconstruction algorithms are developed.

The comparison between the momentum resolution offered by gaseous and solid-state tracking can be studied via analytical algorithms that take point resolution and material burden and orientation into account in calculating the expected momentum resolution for a given detector design proposal (in this paper, I make use of the LCDTRK [5] utility). First, though, it is helpful to present a few basic scaling laws.

For solenoidal detector geometry, charged-particle momentum is determined through the measurement of the sagitta of the projection of helical charged-particle tracks into the plane transverse to the beam axis. The sagitta, in turn, is the difference of the position of the trajectory at the middle of the radial lever arm to the segment connecting the trajectory at its minimal and maximal radial extent. Thus, the accurate measurement of charged particle momentum requires an accurate assessment of the trajectory's location at these three points (minial, maximal, and median radial extent).

An obvious figure of merit can be formed that expresses the accuracy of the measurement of the trajectory in a region of space. If the measurement is performed with N tracking layers, each with a space-point resolution of $\sigma$ μm, the accuracy of the combined measurement will be $\sigma/\sqrt{N}$. A sensible figure of merit is thus $\eta = \sigma/\sqrt{n}$ (μm-$\sqrt{cm}$), where n is the radial readout density in readout segments per centimeter. While, particularly for the case of gaseous tracking, the measurements tend to be distributed within a range of radii, the figure of merit $\eta$ nonetheless provides a meaningful parameter with which to compare tracking technologies.

Roughly speaking, TPC readout tends to have a radial segmentation of about 0.6 cm, and a point resolution of approximately 100 μm, providing a figure of merit $\eta_{TPC} \approx 75$ μm-$\sqrt{cm}$. Solid-state trackers under consideration for the ILC detector, on the other hand, tend to have layers situated with approximately 20 cm separation, with an expected point resolution of 7 μm, leading to a figure of merit of $\eta_{SS} \approx 30$ μm-$\sqrt{cm}$. Thus, as currently applied, solid-state tracking has a finer intrinsic precision than gaseous tracking. In addition, the information provided by solid-state tracking is relatively localized, allowing better advantage to be taken of the full radial extent of the tracker.

It's also interesting to consider a related figure of merit $\eta' = \sqrt{x_0}\eta$ (μm), where $x_0$ is fractional radiation length per cm for the given tracking technology – roughly 0.01% per cm for gaseous tracking, and 0.04% per cm for solid-state

**ALCPG1312**

tracking with layers of 0.8% $X_0$ spaced by 20cm. This figure of merit is relevant for the tracking of lower-momentum particles, for which scattering begins to dominate the accuracy of the momentum determination. For gaseous and solid-state tracking, this yields $\eta'_{TPC} \approx 0.8$ μm and $\eta'_{SS} \approx 0.6$ μm, respectively. Thus, as currently implemented, solid-state tracking provides somewhat more information per radiation length of scattering material than does gaseous tracking, someone contrary to common knowledge. However, signal-to noise considerations force sold-state tracking layers to be relatively thick, and so in order to provide enough layers for efficient pattern recognition, solid-state tracking detectors tend to have a larger overall material burden than gaseous trackers.

Because solid-state tracking is typically considered for compact, high-field detector designs, it tends to rely more heavily on excellent space-point resolution than does TPC-based tracking. However, for the implementation of solid-state and gaseous tracking within identical tracking volumes, both the high-momentum curvature resolution and the calibration accuracy are characterized by the figure of merit η; one sees that the more demanding requirement for solid-state tracker calibration is due to the fact that the tracking is expected to measure the curvature more accurately – for similar curvature resolution performance, the calibration requirement is the same for gaseous and solid-state tracking.

In addition to TPC-based and solid-state tracking designs, it may be worth considering hybrid tracking systems, combining the robust pattern recognition of TPC tracking with the precise point resolution of solid-state tracking. In fact, both existing TPC-based tracking designs (LDC, GLD) employ some degree of microstip tracking.

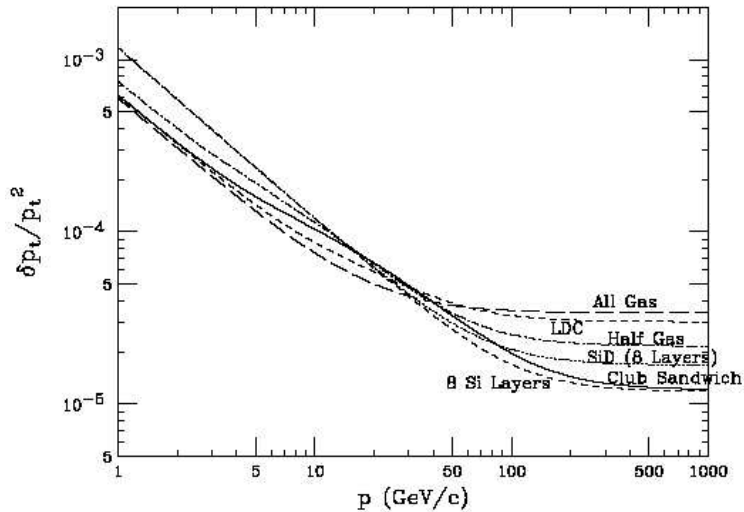

Figure 1: Tracking performance for several different implementations of the LDC tracking volume.

The tradeoffs between the use of TPC-based and solid-state tracking can be evaluated in Figure 1, which shows the curvature resolution at cos θ = 0 as a function of momentum for various ways of instrumenting the LDC tracking volume (assuming a five-layer pixel vertex detector with a radius of 6 cm). For the "Half-Gas" approach, silicon microstrip tracking is used out to half the total tracking radius, with a TPC implement for larger radii. For the "Club Sandwich" approach, a TPC fills the entire tracking volume, except for several layers of silicon microstrips at ½ the full tracking radius and again at the full tracking radius. The performance of the SiD concept, with 8 rather than 5 tracking layers, is also included. As suggested, the intrinsic accuracy of microstrip tracking leads to superior momentum resolution for energetic tracks – a regime that is coming to be understood as critical for the fullest exploitation of the potential of the ILC facility.

## References

[1] T. Barklow, "Physics Impact of Detector Performance", talk given at the 2005 International Linear Collider Workshop, March 18-22, 2005, Stanford, California.

[2] H. Yang, K. Riles "Impact of Tracker Design on Higgs/SUSY Measurement", talk given at the 2005 International Linear Collider Workshop, March 18-22, 2005, Stanford, California.

[3] S. Gerbode *et al.*, "Selectron Mass Reconstruction and the Resolution of the Linear Collider Detector", submitted to the proceedings of the 2005 International Linear Collider Workshop, March 18-22, 2005, Stanford, California; hep-ex/0507053.

[4] C. J. S. Damerell *et al.,* "Ideas for the NLC Detector", published in the Proceedings of the 1996 DPF/DPB Summer Study on New Directions in High-Energy Physics, Snowmass, CO, July 1996.

[5] B. A. Schumm, LCDTRK.F, http://ww.slac.stanford.edu/~schumm/lcdtrk20011204.tar.gz.